\documentclass[aps,eqsecnum,preprint]{revtex4}
\usepackage{graphicx,amsmath}
\begin{document}
\title{Quantum description of classical apparatus:
Zeno effect and decoherence}
\author{S.~A. Gurvitz}
\affiliation{Weizmann Institute of Science, Department of Particle  
Physics, 76100 Rehovot, Israel}
\date{\today}
\begin{abstract}
We study the measurement process by treating classical 
detectors entirely quantum mechanically. As a generic model
we use a point-contact detector coupled to an electron
in a quantum dot and tunneling into the continuum. Transition
to the classical description and the mechanism of decoherence 
are investigated. We concentrate on the influence of the 
measurement on the electron decay rate to the continuum. We
demonstrate that the Zeno (or the anti-Zeno) effect requires
a nonuniform density of states in the continuum. In this case
we show that the anti-Zeno effect relates only to the average 
decay rate, whereas for sufficiently small time the Zeno effect
always takes place. 
We discuss the experimental consequences of our results and 
the role of the projection postulate in a measurement process.
\end{abstract}
\maketitle
\section{Introduction}
The description of a measurement process has been a topic debated from 
the early development of quantum mechanics \cite{neu,whee}. 
Nevertheless, the understanding of quantum-mechanical measurements
has not been achieved yet. The main problem is still the nature of 
the projection postulate \cite{neu}, according to which the 
wave-function of the observed system is projected onto an 
eigenstate of the observable under consideration. During recent years 
the measurement problem received a great deal of attention  
due to exiting opportunities offered by developments in experimental 
techniques of optics and mesoscopic structures. 
The problem also has close connections to the rapidly growing fields 
of quantum cryptography and quantum computing. 

One of the most striking problems, directly related to the 
projection postulate, is the so-called ``Zeno paradox'' (or ``Zeno effect''),
which suggests that frequent or continuous observations can inhibit 
(or slow down) the decay an unstable quantum system \cite{zeno}.
During the last two decades the Zeno effect
has become a topic of great interest. It has been discussed in the areas 
of radioactive decay \cite{panov1}, polarized light \cite{r1},
physics of atoms \cite{r2,r3}, neutron physics \cite{r4}, 
quantum optics \cite{r5}, mesoscopic physics \cite{gur1,hacken}
and even in cognitive science \cite{atm}.
Recently, it was proposed that under some conditions repeated
observations could accelerate the average transition rate of
a quantum system, so called the anti-Zeno effect \cite{gur1,r6,az1}.
This effect has been further analyzed in Refs.~\cite{kof,eg1,facchi,evers}.

The Zeno paradox was originally introduced as an effect of 
continuous observation of an unstable
state. Consider for instance a particle localized initially in a potential well,
which decays to the continuum via tunneling through the barrier,
Fig.~1. It is well-known that the probability of finding the
particle inside the well (the probability of survival) drops down exponentially,
$P_0(t)=e^{-\Gamma t}$.
For small $t$, however, $P_0(t)=1-at^2$ \cite{zeno,per}. Indeed, the probability
of survival is
\begin{equation}
  P_0(t)=|\langle \Phi_0|e^{-iHt/\hbar}|\Phi_0\rangle |^2=1
  -(\Delta H)^2t^2/\hbar^2+\cdots
  \label{a1}
\end{equation}
where $(\Delta H)^2=\langle \Phi_0|H^2|\Phi_0\rangle
-(\langle \Phi_0|H|\Phi_0\rangle )^2$.
\vskip1cm
\begin{minipage}{13cm}
\begin{center}
\includegraphics[width=8cm]{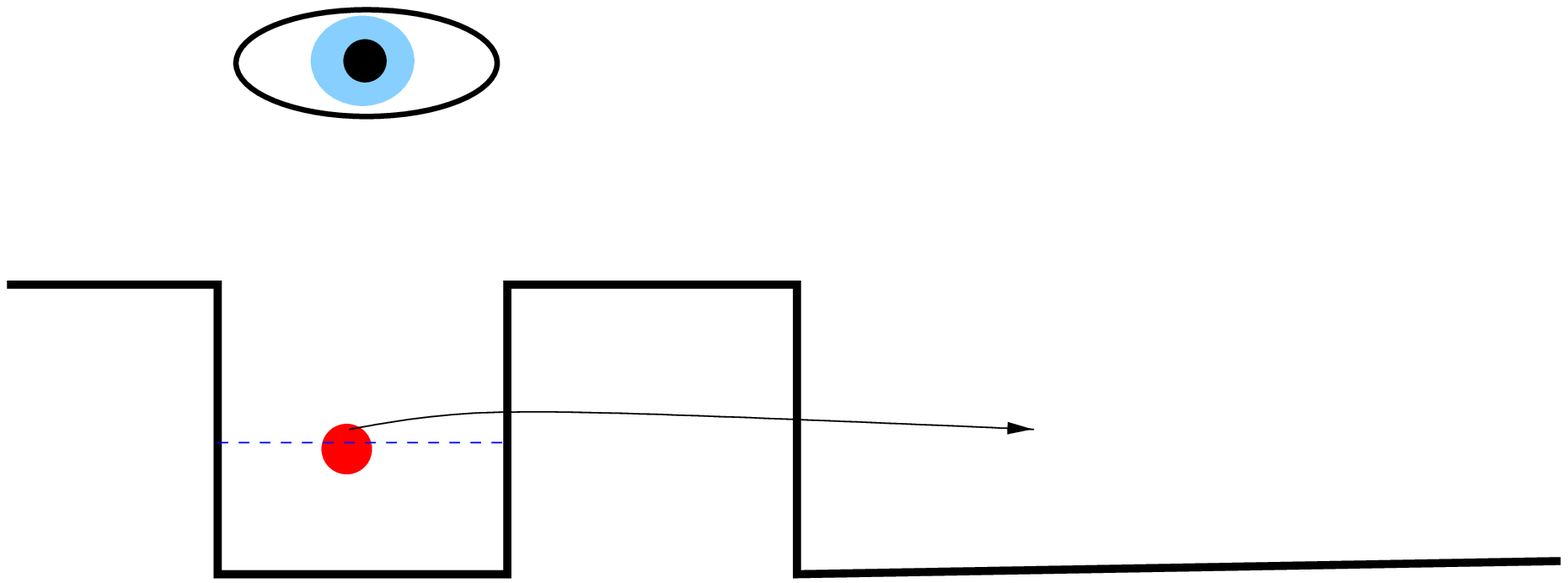}
\end{center}
{\begin{small}
{\bf Fig.~1.} Continuous monitoring of an unstable system decay to
continuum.
 \end{small}}
\end{minipage} \\ \\ 

Let us assume that a particle inside the well is continuously
monitored  (by the ``eye'', shown in Fig.~1). This process can be
viewed as $n$ discrete measurements, where each one takes some
small measurement time $\Delta t$. Then after the first
measurement one finds the particle inside the well with a probability
$P_0(\Delta t)=1-a(\Delta t)^2$. According to the projection
postulate \cite{neu}, the measurement projects the system into the
state which is actually observed. As a result the system continues its evolution 
with the new initial conditions. Hence, after $n$
consecutive measurements the probability of finding the system undecayed
at time $t=n \Delta t$ is
\begin{equation}
  P_0(n\Delta t)=[1-a(\Delta t)^2]^n\, ,
   \label{a2}
 \end{equation}
Taking the limit of continuous monitoring,
$\Delta t\to 0$ and $n\to\infty$, where $t$=const, 
one finds 
\begin{equation}
P_0(t)=[1-a(\Delta t)^2]^{t/\Delta t}\simeq 1-a(\Delta t)t~\to 1~~~~
{\mbox {for}}~~~ \Delta t\to 0. 
  \label{a3}
\end{equation}
Therefore the continuously observed system cannot decay.  

The Zeno paradox in quantum mechanics is still not so famous as   
the  EPR or the Schr\"odinger cat paradoxes. 
Yet, the Zeno paradox represents a real dynamical effect of
the projection postulate and not only an interpretation problem 
of Quantum Mechanics with no experimental consequences \cite{bal}.
For a proper understanding of the Zeno paradox and therefore a role
of the projection postulate in quantum mechanics, it is absolutely
necessary to include the measurement device in the Schr\"odinger
equation for the entire system. In this case quantum-mechanical description
of the measurement device would allow us to study thoroughly
the measurement process without explicit use of the projection postulate. 
The main difficulty with such an approach,
however, is that the measurement device is a macroscopic 
system and therefore its quantum mechanical analysis is very complicated. 
For this reason one would expect that mesoscopic systems,
which are between the microscopic and macroscopic scales,
would be very useful for this investigation \cite{imry}. 

In the following we concentrate on measurements of quantum dots
in a two-dimensional electron gas. As a generic example of
the measurement device (detector) we consider
a point contact (tunneling junction) created electrostatically by
two electrodes. This junction separates two reservoirs, Fig.~2a, which
are filled up to their Fermi levels $\mu_L$ and $\mu_R$, respectively,
with $\mu_L > \mu_R$. As a result, a  macroscopic current $I$
flows through the point contact, as shown  schematically in Fig.~2b,
where the point contact is represented by a potential barrier.
\vskip1cm
\begin{minipage}{13cm}
\begin{center}
\includegraphics[width=10cm]{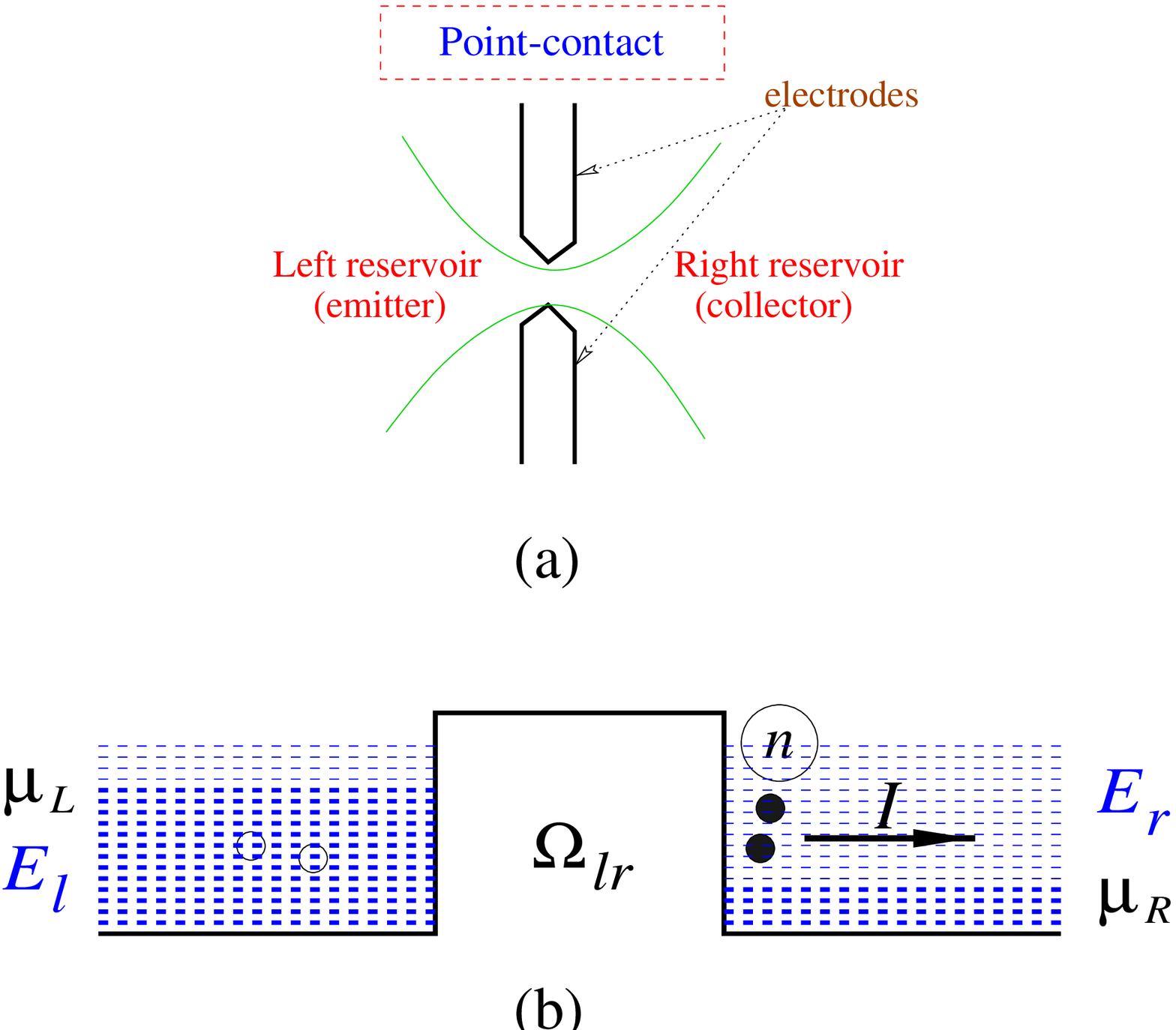}
\end{center}
{\begin{small}
{\bf Fig.~2.} (a) Point-contact in a two-dimensional electron gas
and (b) its schematic representation. $\Omega_{lr}$ is the coupling
between the levels $E_l$ and $E_r$ in the left and the right reservoirs, and 
$n$ denotes the number of electrons arriving the right reservoir by time $t$.
 \end{small}}
\end{minipage} \\ \\ 

The electron current $I$ flowing though the point-contact is
very sensitive to nearby electrostatic fields that modulate the size
of the opening. Thus, one can use the point contact as a detector to monitor
the charging of a quantum dot. For instance, it can be used for monitoring
a single electron in a coupled-dot (electrostatic q-bit), Fig.~3.
Indeed, if the electron occupies
the lower dot, located far away from the point contact (Fig.~3a),
its electric field does not affect the conductivity of
the point contact. However, if the electron occupies the upper
dot,  close to the point-contact, its electrostatic field diminishes
the conductivity of the point contact. As a result the point-contact current
decreases, $I_2<I_1$ (Fig.~3b). Thus, by observing a variation 
of the detector current one can monitor the electron's
transitions between the dots.

In a similar way, by using the point-contact detector 
one can investigate the influence of measurement
on the decay of an unstable system. An appropriate setup 
is shown in Fig.~4. Here the point-contact detector
is placed near the quantum dot, opening into the continuum. Again,
the detector current increases when the electron
leaves the quantum dot.
\vskip1cm
\begin{minipage}{13cm}
\begin{center}
\includegraphics[width=8cm]{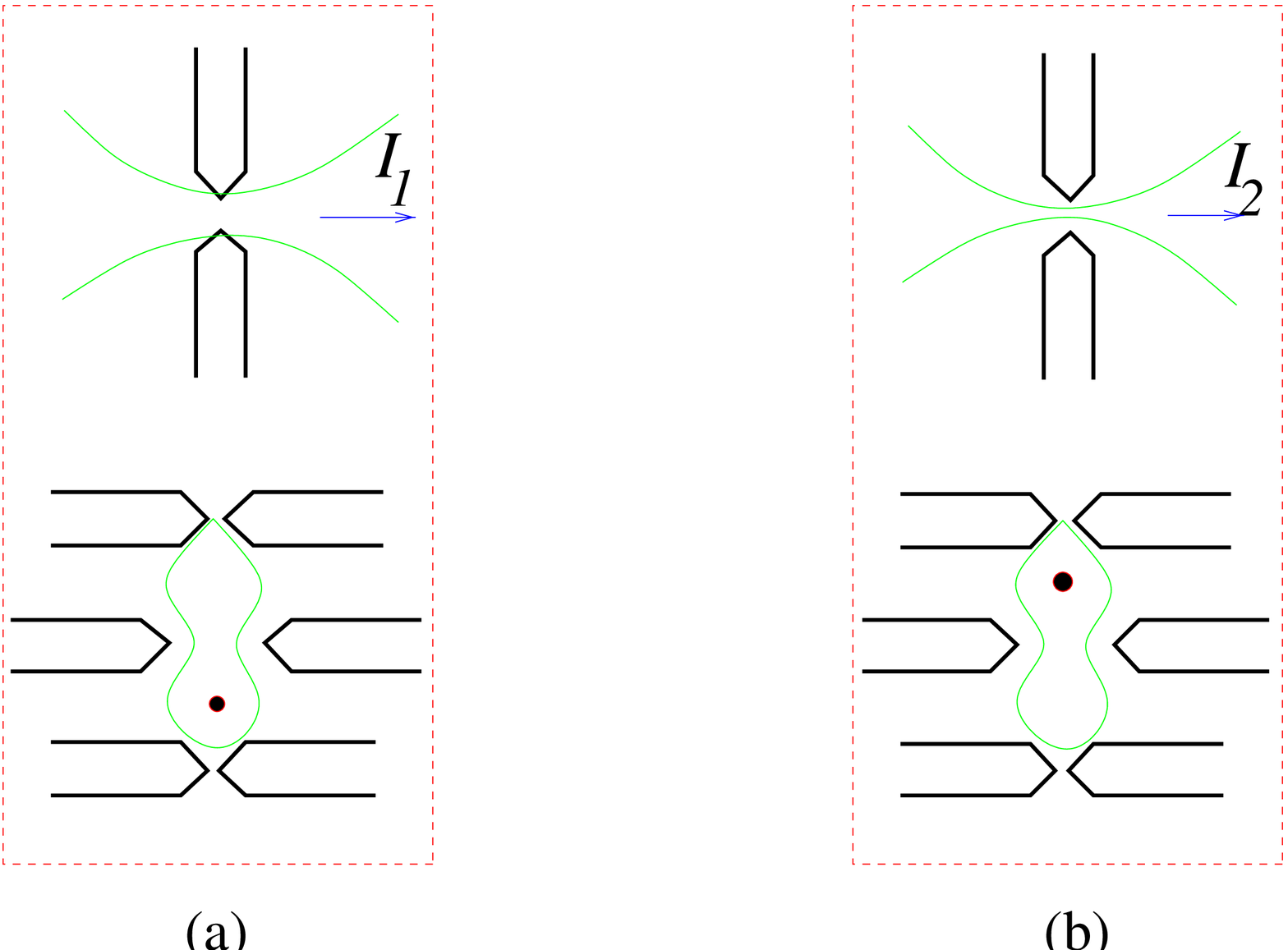}
\end{center}
{\begin{small}
{\bf Fig.~3.} Point-contact as an electrostatic detector of
an electron in a coupled quantum dot. The detector current decreases
upon transition of the electron from the lower to the upper dot. 
 \end{small}}
\end{minipage} \\ \\ 
It should be pointed out that although a quantum point contact
has successfully been used as a ``Which Path Detector''
in different types of experiments (for instance
in measurements of quantum interference \cite{Buks,sprin})
measurements of a ``single electron'' using the
point-contact detector, as in Figs.~3 and 4,
have not been achieved. Nevertheless, the rapid progress of
nano-technology should make such measurements feasible in the near future.

\section{Quantum description of the detector}

Let us consider the point-contact detector and 
the measured electron (Figs.~3,4) as one
quantum system described by the Schr\"odinger equation.
The point-contact detector, however, represents  a macroscopic
system and therefore it should exhibit classical behavior.
This is an essential condition for a ``measurement'' device.  
Now we demonstrate how such a classicality of the
point-contact detector emerges from the Schr\"odinger equation.
\vskip1cm
\begin{minipage}{13cm}
\begin{center}
\includegraphics[width=9cm]{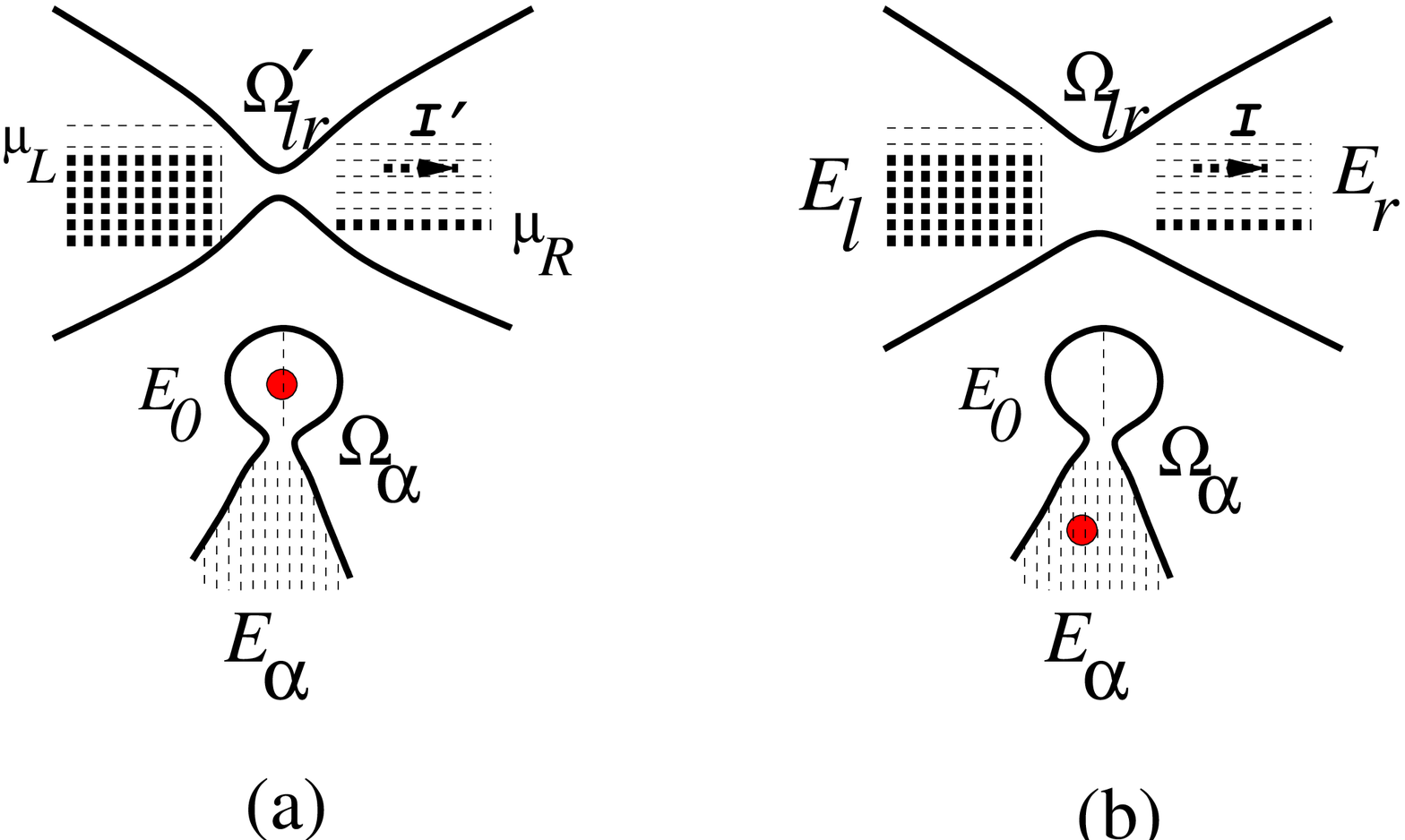}
\end{center}
{\begin{small}
{\bf Fig.~4.} Point-contact monitors of the electron  
in a quantum dot. The detector current increases 
whenever the electron escapes to the continuum.
Here, $\Omega_{lr}$ denotes
the coupling between the left and right reservoirs,
and  $\Omega_\alpha$ is the coupling between the dot and the continuum. 
 \end{small}}
\end{minipage} \\ \\ 

Consider first the point-contact alone, Fig.~2a, 
as described by the following tunneling Hamiltonian
\begin{equation}
{\cal H}_{PC}=\sum_l E_la_l^\dagger a_l+\sum_r E_ra_r^\dagger a_r
+\sum_{l,r}\Omega_{lr}(a_l^\dagger a_r+ a_r^\dagger a_l )\, ,
\label{b1}
\end{equation}
where  $a_l^\dagger (a_l)$ and $a_r^\dagger  (a_r)$ are the creation
(annihilation) operators in the left and the right 
reservoirs, respectively, and $\Omega_{lr}$ is the hopping amplitude 
between the states $E_l$ and $E_r$ in the right and the left reservoirs.  
(We choose the gauge where $\Omega_{lr}$ is real).

We assume that all the levels in the emitter (left) and the collector (right)
are initially filled up to their Fermi 
energies $\mu_L$ and $\mu_R$ respectively. We shall call 
this the ``vacuum'' state, $|0\rangle$.
The Hamiltonian Eq.~(\ref{b1}) requires the vacuum state $|0\rangle$ to decay 
exponentially to a continuum state 
having the form: $a_{r}^{\dagger}a_{l}|0\rangle$ with an electron
in the collector reservoir 
and a hole in the emitter reservoir; 
$a_{r}^{\dagger}a_{r'}^{\dagger}a_{l}a_{l'}|0\rangle$ 
with two electrons in the collector reservoir and two holes in 
the emitter reservoir, and so on. In order to treat such a system
one usually uses the Keldysh non-equilibrium Green's function technique \cite{kel}.
Here we use a different, simpler and more transparent 
technique developed by us in Ref.~\cite{gp}. It consists of reduction
of the Schr\"odinger equation to Bloch-type rate equations
for the density matrix obtained by integrating over the reservoir states. 
Such a procedure is described below, and can be carried out
in the strong nonequilibrium limit without any stochastic assumptions. 

Let us consider the many-body wave function describing
the point-contact detector. It can be written 
in the occupation number representation as 
\begin{equation}
|\Psi (t)\rangle = \left [ b_0(t) + \sum_{l,r} b_{lr}(t)a_r^{\dagger}a_l
+\sum_{l<l',r<r'} b_{ll'rr'}(t)a_r^{\dagger}a_{r'}^{\dagger}a_la_{l'}
+\cdots\right ]|0\rangle, 
\label{b2}
\end{equation}
where $b(t)$ are the time-dependent probability amplitudes for
finding the system in the corresponding states given the initial 
condition $b_0(0)=1$, with all the other $b(0)$'s being zeros.
In order to find these amplitudes we 
substitute Eq.~(\ref{b2}) into the Schr\"odinger equation 
$i\,|\dot\Psi (t)\rangle ={\cal H}_{PC}|\Psi (t)\rangle$. As
a result we obtain an infinite system of linear differential equations
for the amplitudes $b(t)$, which completely determines the 
quantum behavior of the point contact. It is useful to use 
the Laplace transform, 
$\tilde{b}(E)=\int_0^{\infty}e^{iEt}b(t)dt$, so that these differential
equations become the following algebraic coupled equations
for the amplitudes $\tilde{b}(E)$ 
\begin{subequations}
\label{b3}
\begin{eqnarray}
& &E \tilde{b}_{0}(E) - \sum_{l,r} \Omega_{lr}\tilde{b}_{lr}(E)=i
\label{b3a}\\
&(&E + E_{l} - E_r) \tilde{b}_{lr}(E) - \Omega_{lr}\tilde{b}_0(E) 
-\sum_{l',r'}\Omega_{l'r'}\tilde{b}_{ll'rr'}(E)=0
\label{b3b}\\
&(&E + E_{l}+E_{l'} - E_r-E_{r'}) \tilde{b}_{ll'rr'}(E) 
- \Omega_{l'r'}\tilde{b}_{lr}(E)+\Omega_{lr}\tilde{b}_{l'r'}(E)
\nonumber\\
&&~~~~~~~~~~~~~~~~~~~~~~~~~~~~~~~~~~~~~~~~~~~~~~~~~~~~~
-\sum_{l'',r''}\Omega_{l''r''}\tilde{b}_{ll'l''rr'r''}(E)=0
\label{b3c}\\
&&~~~~~~~~~~~~~~~~~~~~~~~~~~~~~~~~~~~~~~\cdots\
\nonumber
\end{eqnarray}
\end{subequations}

Eqs. (\ref{b3}) can be substantially simplified by replacing 
the amplitude $\tilde b$ in 
the term $\sum\Omega\tilde b$ of each of the equations  by 
its expression obtained from the subsequent equation \cite{gp}.  
For example,   
substituting $\tilde{b}_{lr}(E)$ from Eq.~(\ref{b3b}) into Eq.~(\ref{b3a}), 
one obtains
\begin{equation}
\left [ E - \sum_{l,r}\frac{\Omega^2}{E + E_{l} - E_r}
    \right ] \tilde{b}_{0}(E) - \sum_{ll',rr'}
    \frac{\Omega^2}{E + E_{l} - E_r}\tilde{b}_{ll'rr'}(E)=i,
\label{ap5}
\end{equation}
where we assumed that the hopping amplitudes 
are weakly dependent functions on the energies
$\Omega_{lr}\equiv\Omega (E_l,E_r)=\Omega$.
Since the states in the reservoirs are very dense (essential a continuum), 
one can replace the sums over $l$ and $r$ by integrals, for instance  
$\sum_{l,r}\;\rightarrow\;\int \rho_{L}(E_{l})\rho_{R}(E_{r})\,dE_{l}dE_r\:$,
where $\rho_{L,R}$ are the densities of states in the emitter and collector. 
Then the first sum in Eq.~(\ref{ap5}) becomes an
integral which can be split into the sum of its singular and principal value 
parts. The singular part yields $i\pi\Omega^2\rho_L\rho_R V_d$,
and the principal part is merely absorbed into a
redefinition of the energy levels. The second sum in Eq.~({\ref{ap5}) 
can be neglected. Indeed, by replacing 
$\tilde{b}_{ll'rr'}(E)\equiv \tilde{b} (E,E_l,E_{l'},E_r,E_{r'})$ and 
the sums by the integrals we find that the integrand  
has the poles on the same sides of the integration 
contours. It means that the corresponding integral vanishes for
infinite integration limits. This corresponds to strongly
non-equilibrium limit, $V_d/\Omega^2\rho\to\infty$.  

Applying analogous considerations to the other equations of the
system (\ref{b3}), we finally arrive at the following set of equations: 
\begin{subequations}
\label{ap6}
\begin{eqnarray}
&& (E + iD/2) \tilde{b}_{0}(E)=i
\label{ap6a}\\
&& (E + E_{l} - E_r + iD/2) \tilde{b}_{lr}(E)
      - \Omega\tilde{b}_{0}(E)=0
\label{ap6b}\\ 
&& (E + E_{l}+ E_{l'} - E_{r} - E_{r'} + iD/2) \tilde{b}_{ll'rr'}(E) -
      \Omega\tilde{b}_{lr}(E)+\Omega \tilde{b}_{l'r'}(E)=0,
\label{ap6c}\\
& &~~~~~~~~~~~~~~~~~~~~~~~~~~~~~~~~~~~~\cdots 
\nonumber
\end{eqnarray}
\end{subequations}
where $D=2\pi\Omega^2\rho_L\rho_R V_d$. 

\subsection{Rate equations} 

Eqs.~(\ref{ap6}) can be transformed to differential equations
for the reduced density matrix $\sigma^{(nn')}(t)$
of electrons in the right reservoir (collector). This density matrix is directly
related to the amplitudes $\tilde b(t)$. For instance
the diagonal density-matrix elements 
\begin{center}
\mbox{\includegraphics[width=10cm]{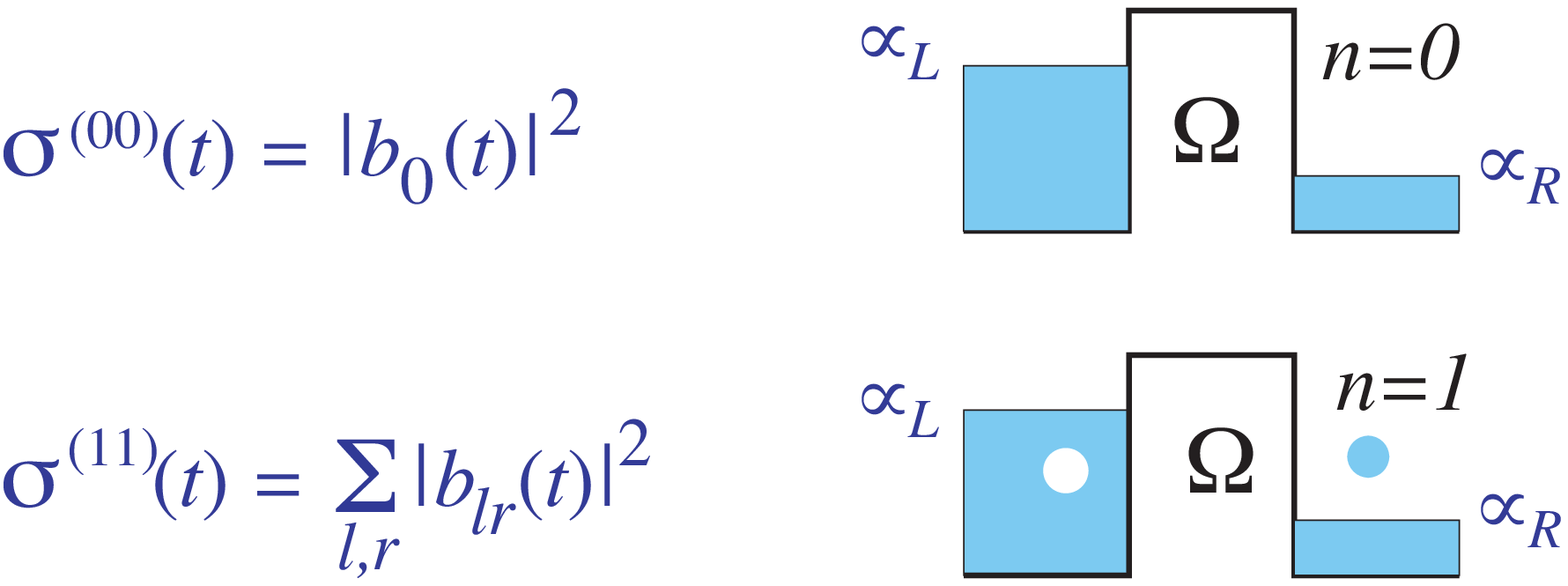}}
\end{center}
\hspace*{35mm}$\cdots\cdots\cdots$\\ 
are the probabilities of finding $0$, $1$, etc.~electrons in the
collector. The corresponding off-diagonal matrix elements (coherences)
\begin{equation}
\sigma^{(01)}(t)=\sum_{l,r}b_0(t)b^*_{lr}(t),~~~\sigma^{(12)}(t)=
\sum_{l<l'\atop r<r'}b_{lr}(t)b^*_{ll',rr'}(t),\ldots
\label{app8}
\end{equation}  
have no classical equivalent and describe electrons in the linear
superposition of the states in different reservoirs.

Let us rewrite $\sigma^{(nn')}(t)$ in terms of the  
amplitudes $\tilde b(E)$ via the inverse Laplace transform 
\begin{equation}
\sigma^{(n,n')}(t)=
\sum_{l\ldots , r\ldots}
\int \frac{dEdE'}{4\pi^2}\tilde b_{\underbrace{l\cdots}_{n}
  \underbrace {r\cdots}_{n}}(E)
\tilde b^*_{\underbrace{l\cdots}_{n'}
  \underbrace {r\cdots}_{n'}}(E')e^{i(E'-E)t}
\label{ap9}
\end{equation}
Using Eq.~(\ref{ap9}) one can transform  Eqs.~(\ref{ap6}) directly into
equations for $\sigma^{(n,n')}(t)$ (c.f. \cite{gur1,gp}).
For instance, consider Eq.~(\ref{ap6b}) for the amplitude $\tilde b_{lr}(E)$. 
Multiplying this equation by $\tilde b_{lr}^*(E')$ and subtracting the
same equation for $\tilde b_{lr}^*(E')$ multiplied by
$\tilde b_{lr}(E)$ we obtain 
\begin{equation}
(E'-E-iD)\tilde b_{lr}(E)\tilde b^*_{lr}(E')-\Omega
\left [\tilde b_{lr}(E)\tilde b^*_0(E')
-\tilde b^*_{lr}(E')\tilde b_0(E)\right ]=0\, .
\label{app9}
\end{equation}
Substituting Eq.~(\ref{app9}) into Eq.~(\ref{ap9})
we find
\begin{equation}
\dot{\sigma}^{(1,1)}(t)=-D\sigma^{(1,1)}(t)-i\Omega 
\left [\sigma^{0,1)}(t)-\sigma^{(1,0)}(t)\right ]
\label{app10}
\end{equation}

A similar procedure can be performed for all other equations
(\ref{ap6}). As a result we arrive at the following system of
equations for the density matrix $\sigma^{(n,n')}$
\begin{subequations}
\label{ap10}
\begin{eqnarray}  
&&\dot{\sigma}^{(n,n)}(t)=-D\sigma^{(n,n)}(t)-i\Omega 
\left [\sigma^{(n-1,n)}(t)-\sigma^{(n,n-1)}(t)\right ]
\label{a10a}\\
&&\sigma^{(n-1,n)}(t)=i(D/\Omega )\sigma^{(n-1,n-1)}(t)
\label{a10b}
\end{eqnarray}
\end{subequations}
Substituting Eq.~(\ref{a10b}) into Eq.~(\ref{a10a}) we find
a linear differential equation for the probabilities only
\begin{equation}
\frac{d}{dt}\sigma^{(n,n)}(t) = -D\sigma^{(n,n)}(t)
+D\sigma^{(n-1,n-1)}(t)\, ,
\label{aa11}
\end{equation}
where
$D=2\pi\Omega^2\rho_L\rho_R(\mu_L-\mu_R)$ is the rate of electrons
arriving in the right reservoir.

Equation (\ref{aa11})is a classical rate equation. On the other
hand it was obtained from the Schr\"odinger equation entirely in the
framework of Quantum Mechanics. 
No Markov approximations have been used in its derivation.
It is important to note that this classical limit of the
Schr\"odinger equation does not require the vanishing of
the nondiagonal density-matrix elements (coherences), 
as follows from Eq.~(\ref{a10b}). The transition from quantum to classical
description is provided by a {\em decoupling} of coherences from
probabilities in the equation of motion, Eq.~(\ref{aa11}).

Equation (\ref{aa11}) can easily be solved for the initial condition 
$\sigma^{(n,n)}(0)=\delta_{n0}$. One finds the Poisson distribution
\begin{equation}
\sigma^{(n,n)}(t)=\frac{(D\ t)^n}{n!}e^{-D\ t}\simeq
\frac{1}{\sqrt{2\pi Dt}}\exp\left [-\frac{(Dt-n)^2}{2Dt}\right ]
\label{b4}
\end{equation}
Thus the average electric current flowing into the
right reservoir is $I=e<n>/t=eD$. 

\section{Measurement of the decay time}

Consider now the measurement of the Zeno effect for decay into
the continuum using a point-contact detector, as in Fig.~4.  
The tunneling Hamiltonian describing the entire system
consists of three parts \cite{eg,eg1}:
\begin{equation}
{\cal H}={\cal H}_{PC}+{\cal H}_{QD}+{\cal H}_{int}.
\label{c0}
\end{equation}  
The first term
describes tunneling through the point-contact detector, Eq.~(\ref{b1}). 
The second term describes the quantum dot coupled to the continuum. 
\begin{equation}
{\cal H}_{QD} = E_0 c_0^{\dagger}c_{0}
+\sum_\alpha E_\alpha c^\dagger_\alpha c_\alpha
+\sum_\alpha \Omega_\alpha
(c_\alpha^\dagger c_0+c_0^\dagger c_\alpha)
\label{c1}
\end{equation}
Here the operators $c_0^\dagger (c_0)$ and $c^\dagger_\alpha
(c_\alpha)$ create (annihilate) an electron inside the dot or
in the continuum, respectively, and $\Omega_\alpha$ is the
corresponding coupling between these states.

The last term describes the interaction of the detector
with the electron in the quantum dot,
\begin{equation}
{\cal H}_{int}=-\sum_{l,r}\delta\Omega_{lr}c_0^{\dagger} 
c_0(a^{\dagger}_la_r +a^{\dagger}_ra_l).
\label{c2}
\end{equation}
This term modulates the detector current.  
That is whenever the electron occupies the dot,
i.e. $c_0^{\dagger}c_{0}\to 1$, the coupling
between the states $E_{l,r}$ of the detector decreases
$\Omega'_{lr}=\Omega_{lr}-\delta\Omega_{lr}<\Omega_{lr}$.
In this case the detector current becomes $I'=eD'$, where
$D'=2\pi(\Omega-\delta\Omega_{lr})^2\rho_L\rho_R(\mu_L-\mu_R)$.

\subsection{Dynamics of an unstable system} 

Consider first the decay of an electron into the continuum with no interaction    
with the point-contact detector, $\delta\Omega_{lr}=0$. In this case
the electron evolution is determined only by 
${\cal H}_{QD}$,  Eq.~(\ref{c1}). This Hamiltonian is essentially
equivalent to that of the Lee model. The latter reproduces an exponential decay
for homogeneous reservoirs and a constant
coupling $\Omega_\alpha$, without any Markovian approximations \cite{pfeifer}.
Let us demonstrate this result by solving the time-dependent
Schr\"{o}dinger equation
$i\partial_t|\Psi(t)\rangle={\cal H}_{QD}|\Psi(t)\rangle$.
The electron wave function can be written in the most general way as  
\begin{eqnarray}
|\Psi (t)\rangle = \left [ b_0(t)c^\dagger_0 
+\sum_{\alpha} b_{\alpha}(t)c_{\alpha}^{\dagger} \right ]|0\rangle\ ,
\label{c3}
\end{eqnarray}
where $b_{0,\alpha}(t)$ are the time-dependent probability amplitudes for finding 
the electron inside the dot or in the reservoir 
in the state $E_\alpha$. The initial condition is $b_0(0)=1$
and $b_\alpha (0)=0$. Performing the Laplace transform,
$b(t)\to\tilde b(E)$, we easily find that the Schr\"{o}dinger equation
can be written as
\begin{subequations}
\label{c4}
\begin{eqnarray}
&&(E-E_0)\tilde{b}_0(E)-\sum_{\alpha}\Omega_\alpha \tilde{b}_\alpha = i\ ,
\label{c4a}\\
&&(E-E_\alpha)\tilde b_\alpha (E)-\Omega_\alpha \tilde b_0(E) = 0\ .
\label{c4b} 
\end{eqnarray}
\end{subequations}
In order to solve these equations we replace the amplitude 
$\tilde b_\alpha$ in Eq.~(\ref{c4a}) by its expression obtained from 
Eq.~(\ref{c4b}). One then obtains
\begin{equation}
\left [ E-E_0-\sum_\alpha \frac{\Omega^2_\alpha }{E-E_\alpha} 
\right ]\tilde b_0(E)=i .
\label{c5}
\end{equation}   
Since the states in the reservoir are very dense, one can replace 
the sum over
$\alpha$ by an integral over $E_\alpha$. 
\begin{equation}
\sum_\alpha \frac{\Omega^2_\alpha }{E-E_\alpha} =
\int\frac{\Omega^2 (E_\alpha )
\rho (E_\alpha )}{E-E_\alpha }dE_\alpha \, ,
\label{c6}
\end{equation}   
where $\rho (E_\alpha )$ is the density of states in the reservoir. 
To evaluate this integral, we can split
the integral into its principal and singular parts,  
$-i\delta (E-E_\alpha)$. As a result the original Schr\"odinger equation  
(\ref{c4}) is reduced to  
\begin{subequations}
\label{c7}
\begin{eqnarray}
&&\left [E-E_0-\Delta (E)+i\frac{\Gamma (E)}{2}\right ]\tilde b_0(E)= i\, ,
\label{c7a}\\
&&\left (E-E_\alpha\right )\tilde b_\alpha (E)-\Omega (E_\alpha )
\tilde b_0(E) = 0 ,
\label{c7b}
\end{eqnarray}
\end{subequations}   
where $\Gamma (E)=2\pi \rho(E)\Omega_\alpha^2(E)$ 
and $\Delta (E)$ is the  
energy-shift due to the principal part of the integral.

Let us assume that $\Omega_\alpha^2 (E_\alpha )\rho (E_\alpha )$ 
is weakly dependent on the energy $E_\alpha$. As a result 
the width becomes a constant $\Gamma (E)=\Gamma_0$
and the energy shift $\Delta (E)$ tends to zero. 
Using Eqs.~(\ref{c7}) and the inverse Laplace transform one 
obtains the occupation probabilities of the levels 
$E_0$ and $E_\alpha$ \cite{gur}. Yet, Eqs.~(\ref{c7}) are 
not convenient if we wish to include the effects of a measurement   
(or of an environment) on the electron behavior.
These effects can be determined in a natural way only in terms  
of the density matrix. For this reason  
we transform Eqs.~(\ref{c7}) into  
equations for the density matrix   
$\sigma_{ij}(t)\equiv b_i(t)b^*_j(t)$. The latter is directly related 
to the amplitudes $\tilde b(E)$ via the inverse Laplace transform,
Eq.~(\ref{ap9}). One finds
\begin{subequations}
\label{Reqn1}
\begin{eqnarray}
&&\dot \sigma_{00}(t) = -\Gamma_0\sigma_{00}(t) ,
\label{Reqn1a}\\
&&\dot \sigma_{\alpha\alpha}(t) = i\Omega_\alpha 
(\sigma_{\alpha 0}(t)-\sigma_{0\alpha}(t))
\label{Reqn1b}\\
&&\dot \sigma_{\alpha 0}(t) = i\epsilon_{0\alpha }
\sigma_{\alpha 0}(t)-i\Omega_\alpha \sigma_{00}(t)
-\frac{\Gamma_0}{2}\sigma_{\alpha 0}(t)\ ,
\label{Reqn1c}
\end{eqnarray}
\end{subequations}  
with $\epsilon_{0\alpha }=E_0-E_\alpha$ and $\sigma_{0\alpha}
=\sigma^*_{\alpha 0}$. Here $\sigma_{00}(t)$ and
$\sigma_{\alpha\alpha}(t)$ are the probabilities of finding the
electron in the dot or in the continuum at the level $E_\alpha$,
respectively. The off-diagonal density-matrix elements $\sigma_{\alpha 0}(t)$
(coherences) describe the electron in a linear superposition.
These matrix elements decrease exponentially due to the last
term in Eq.~(\ref{Reqn1c}), generated by decay into the continuum.

Eqs.~(\ref{Reqn1}) represent a generalization of the optical   
Bloch equations describing quantum transitions
between two isolated levels \cite{bloch,gp} to transitions between 
one isolated level and the continuum \cite{eg,eg1,eg2}. In this case the coherence term 
$\sigma_{\alpha 0}$ is coupled to the 
corresponding probability term $\sigma_{00}$, but not with that
of the continuum spectrum $\sigma_{\alpha\alpha}$, as one would expect
for usual optical Bloch equations. 

Solving  Eqs.~(\ref{Reqn1}) we find 
the following expressions for the occupation probabilities, 
$\sigma_{00}$ and $\sigma_{\alpha\alpha}$, of the levels 
$E_0$ and $E_\alpha$, respectively \cite{gur}:
\begin{subequations}
\label{Req1}
\begin{eqnarray}
&&\sigma_{00}(t) = e^{-\Gamma_0 t}\ ,
\label{Req1a}\\[5pt]
&&\sigma_{\alpha \alpha}(t) =\frac{\Omega^2_\alpha}{\displaystyle 
(E_\alpha -E_0)^2+(\Gamma_0/2)^2}\left [1-2\cos [(E_\alpha-E_0)t]\,
e^{-\Gamma_0t/2}+e^{-\Gamma_0 t}\right ]
\label{Req1b}  
\end{eqnarray}
\end{subequations}
Notice that the line shape, 
$P(E_\alpha )\equiv \sigma_{\alpha \alpha}(t\to\infty )\rho$, 
given by Eq.~(\ref{Req1b})
is the standard Lorentzian distribution,
\begin{equation}
P(E_\alpha ) =\frac{\Gamma_0/(2\pi )}{\displaystyle 
(E_\alpha -E_0)^2+(\Gamma_0/2)^2} ~,
\label{loren}
\end{equation}
with the width $\Gamma_0$ corresponding to the inverse 
life-time of the quasi-stationary state, Eq.~(\ref{Req1a}). 

\subsection{Continuous measurement of an unstable system.}

Now we introduce the coupling with the point-contact detector,
$\delta\Omega_{lr}\not =0$. The many-body wave function describing
the entire system can be written in the same way as Eq.~(\ref{b2})
\begin{equation}
|\Psi (t)\rangle = \left [ b_0(t)c_0^\dagger 
+\sum_{l,r} b_{lr}(t)a_r^{\dagger}a_lc_0^\dagger
+\sum_\alpha b_\alpha(t)c_\alpha^\dagger c_0
+\sum_{\alpha ,l,r} b_{\alpha lr}(t)a_r^{\dagger}c_\alpha^{\dagger}a_l
c_0+\cdots\right ]|0\rangle
\label{c8}
\end{equation}
where $b(t)$ are the time-dependent probability amplitudes for
finding the system in the corresponding states, given the initial 
condition $b_0(0)=1$ with all other $b(0)$'s equal to zero.
These amplitudes are obtained from the Schr\"odinger equation
$i\,|\dot\Psi (t)\rangle ={\cal H}|\Psi (t)\rangle$, where
${\cal H}$ is given by Eq.~(\ref{c0}).
Using the same technique as in the previous case, Eqs.~(\ref{aa11}) and (\ref{Reqn1}),
the Schr\"odinger equation for the amplitudes
$b(t)$ is reduced to quantum rate equations for the reduced density
matrix $\sigma_{ij}^{nn}(t)\equiv\sigma_{ij}^{n}(t)$ by integration
over the reservoir states of the detector. One finds \cite{eg,eg1}
\begin{subequations}
\label{Reqn}
\begin{eqnarray}
 \label{Reqna} 
\dot{\sigma}_{00}^{(n)} & = & -(\Gamma+D')\sigma_{00}^{(n)}
+D'\sigma_{00}^{(n-1)}\nonumber\\[5pt]
\label{Reqnb}
\dot{\sigma}_{\alpha\alpha}^{(n)} & = & -D\sigma_{\alpha\alpha}^{(n)}+
D\sigma_{\alpha \alpha}^{(n-1)}+i\Omega_\alpha(\sigma_{0\alpha}^{(n)}
-\sigma_{\alpha 0}^{(n)})\nonumber\\[5pt]
\label{Reqnc}
\dot{\sigma}_{\alpha 0}^{(n)} & = & i(E_0-E_\alpha)\sigma_{\alpha 0}^{(n)}
-i\Omega_\alpha \sigma_{00}^{(n)}-\frac{\Gamma_0+D+D'}{2}\sigma_{\alpha 0}^{(n)}
+\sqrt{DD'}\sigma_{\alpha 0}^{(n-1)}\ ,
\end{eqnarray} 
\end{subequations}
where $\Gamma_0=2\pi\rho(E_0)\Omega_\alpha^2(E_0)$ corresponds to the inverse 
lifetime of the quasi-stationary state, Eq.~(\ref{Req1a}).  
The index $n$ denotes the number of electrons arriving the left
reservoir by time $t$ and the indices $i,j=0,\alpha$ denote the state of the
observed electron. One finds that only the density-matrix elements
diagonal in $n$ enter into the rate equations, similar to Eq.~(\ref{aa11}).
However, we are not integrating over the final states of the escaped
electron. As a result the off-diagonal terms, describing the
superposition of the electron in the dot and in the continuum,
enter the rate equations (c.f.~Eqs.~(\ref{Reqn1})).

The density matrix $\sigma_{ij}^{(n)}(t)$ given by Eqs.~(\ref{Reqn})
describes both the detector
and the escaped electron. Indeed the probability of finding $n$
electrons in the collector, $\sigma^{(n)}(t)$, is obtained by tracing
over the escaped electron variables
\begin{equation}
\sigma^{(n)}(t)=\sigma_{00}^{(n)}(t)+\sum_{\alpha}
\sigma_{\alpha\alpha}^{(n)}(t)
\label{c9}
\end{equation}
The average detector current is therefore  
\begin{equation}
  <\!I(t)\!>\:=e\sum_nn\dot\sigma^{(n)}(t)=eD'\sigma_{00}(t)+eD[1-\sigma_{00}(t)]\, ,
\label{c10}
\end{equation}
where $\sigma_{00}(t)=\sum_n\sigma_{00}^{(n)}(t)$ is the probability 
of finding the electron inside the dot. The latter is obtained by
tracing over the detector variables in the total density matrix:
$\sigma_{ij}(t)=\sum_n\sigma^{(n)}_{ij}(t)$.
One easily finds from Eqs.~(\ref{Reqn}) that
\begin{subequations}
\label{eqn}
\begin{eqnarray}
\dot{\sigma}_{00} & = & -\Gamma_0\sigma_{00}
\label{eqna}\\[5pt]
\dot{\sigma}_{\alpha\alpha} & = & i\Omega_\alpha(\sigma_{\alpha 0}
-\sigma_{0\alpha})
\label{eqnb}\\[5pt]
\dot{\sigma}_{\alpha 0} & = & i(E_0-E_\alpha )\sigma_{\alpha 0}-
i\Omega_\alpha \sigma_{00}-\frac{\Gamma_0+\Gamma_d}{2}\sigma_{\alpha 0}\ ,
\label{eqnc}
\end{eqnarray}
\end{subequations}
where $\Gamma_d=(\sqrt{D}-\sqrt{D'})^2$ is 
the decoherence rate generated by the detector, in addition to the
``intrinsic'' decoherence rate $\Gamma_0$ generated by tunneling.
Here we would like to point out the important distinction between
these two different origins of decoherence. Tunneling
into an infinite continuum is the only intrinsically irreversible process 
encountered in ordinary quantum mechanics. On the other hand $\Gamma_d$ is
related to the ``effective'' irreversibility that occurs when a simple
quantum system is coupled to a macroscopic measurement apparatus,
averaged over unobservable degrees of freedom. 

Equation (\ref{c10}) displays a direct connection between
the averaged detector current and the probability of
finding the electron inside the dot. Its escape to the continuum results in   
an increase of the detector current at $t=1/\Gamma_0$, as shown in
Fig.~5. Therefore the continuous measurement process is completely
described by the rate equations (\ref{Reqn}).

Comparing Eqs.~(\ref{eqn}) with Eqs.~(\ref{Reqn1}) we find that the decay rate
is not modified by the detector. Indeed, the probability of finding
the electron inside the dot drops down with the same exponential,
$\sigma_{00}(t)=\exp (-\Gamma_0t)$, as in the noninteracting case
($\delta\Omega_{lr}=0$). Therefore no Zeno
effect can be observed in the exponential decay of an unstable system.\\
\vskip1cm
\begin{minipage}{13cm}
\begin{center}
\includegraphics[width=6cm]{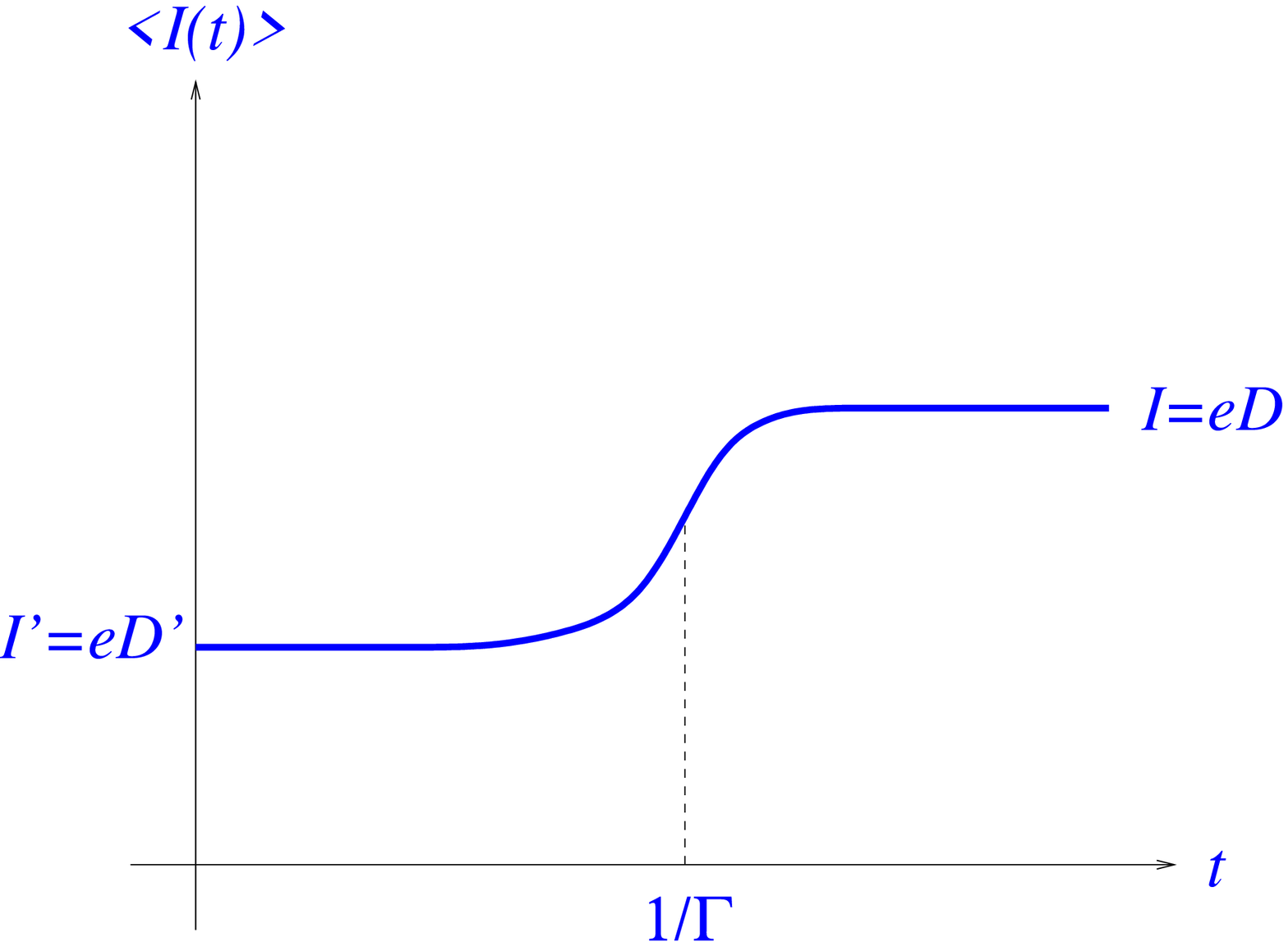}
\end{center}
{\begin{small}
{\bf Fig.~5.} The average detector current as a function of time.
\end{small}}
\end{minipage} \\ \\ 

Nevertheless, the influence of the measurement can be seen in the
last term of Eq.~(\ref{eqnc}), which constitutes an additional decoherence rate
$\Gamma_d$ generated by the detector. This affects the energy spectrum
of the tunneling electron, given by
$P(E_{\alpha})=\sigma_{\alpha\alpha}(t\to\infty)$. Indeed, on
solving Eqs.~(\ref{eqnb}), (\ref{eqnc}) in the limit $t\to\infty$
we obtain
\begin{equation}
P(E_{\alpha})=\frac{\Gamma_0+\Gamma_d}{(E_\alpha-E_0)^2
  +\frac{(\Gamma_0+\Gamma_d)^2}{4}}
\label{c11}
\end{equation}
Comparing with Eq.~(\ref{loren}) one finds that the measurement results in a
broadening of the line width, which becomes $\Gamma_0+\Gamma_d$.

To understand these results, one might think of the following argument. 
Due to the measurement, the energy level $E_0$ suffers an additional
broadening of the order of  $\Gamma_d$. However, this broadening does 
not affect the decay rate of the electron $\Gamma_0$, 
since the exact value of $E_0$ relative to $E_\alpha$ is irrelevant
to the decay process. In contrast, the probability distribution
$P(E_\alpha)$ is affected because it does depend on the position
of $E_0$ relative to $E_\alpha$, as can be seen in Eq.~(\ref{c11}).

Although our result has been proved for a specific detector, 
we expect it to be valid for the general case, provided that the
density of states $\rho$ and the transition amplitude $\Omega_\alpha$ 
for the observed electron vary slowly with
energy. This condition is sufficient to ensure a pure exponential
decay of the state $E_0$ \cite{pfeifer}.  On the other hand, if  
the product $\Omega_\alpha^2\rho (E_\alpha )$ 
depends sharply on energy $E_\alpha$, it yields strong $E$-dependence
of $\Gamma$ and $\Delta$ in  Eqs.~(\ref{c7}).
This would result in a deviation from a pure exponential decay
and consequently to the Zeno effect in the case of continuous measurement.

\section{Nonuniform density of state and Zeno effect.}

Consider the electron escape to the reservoir,
Fig.~4, where the density of the reservoir states $\rho (E_\alpha )$
does depend on the energy. For the definiteness we take
a Lorentzian form of the density of states
\begin{equation}
\rho(E_\alpha )=
\frac{\Gamma_1/2\pi}{(E_\alpha-E_1)^2+\Gamma_1^2/4}
\label{d1}
\end{equation}
One can demonstrate \cite{eg1} that such a system can be mapped onto that
shown in Fig.~6, where the Lorentzian states are represented
by a resonance cavity coupled to the quantum dot and the
reservoir.
\vskip1cm
\begin{minipage}{13cm}
\begin{center}
\includegraphics[width=5cm]{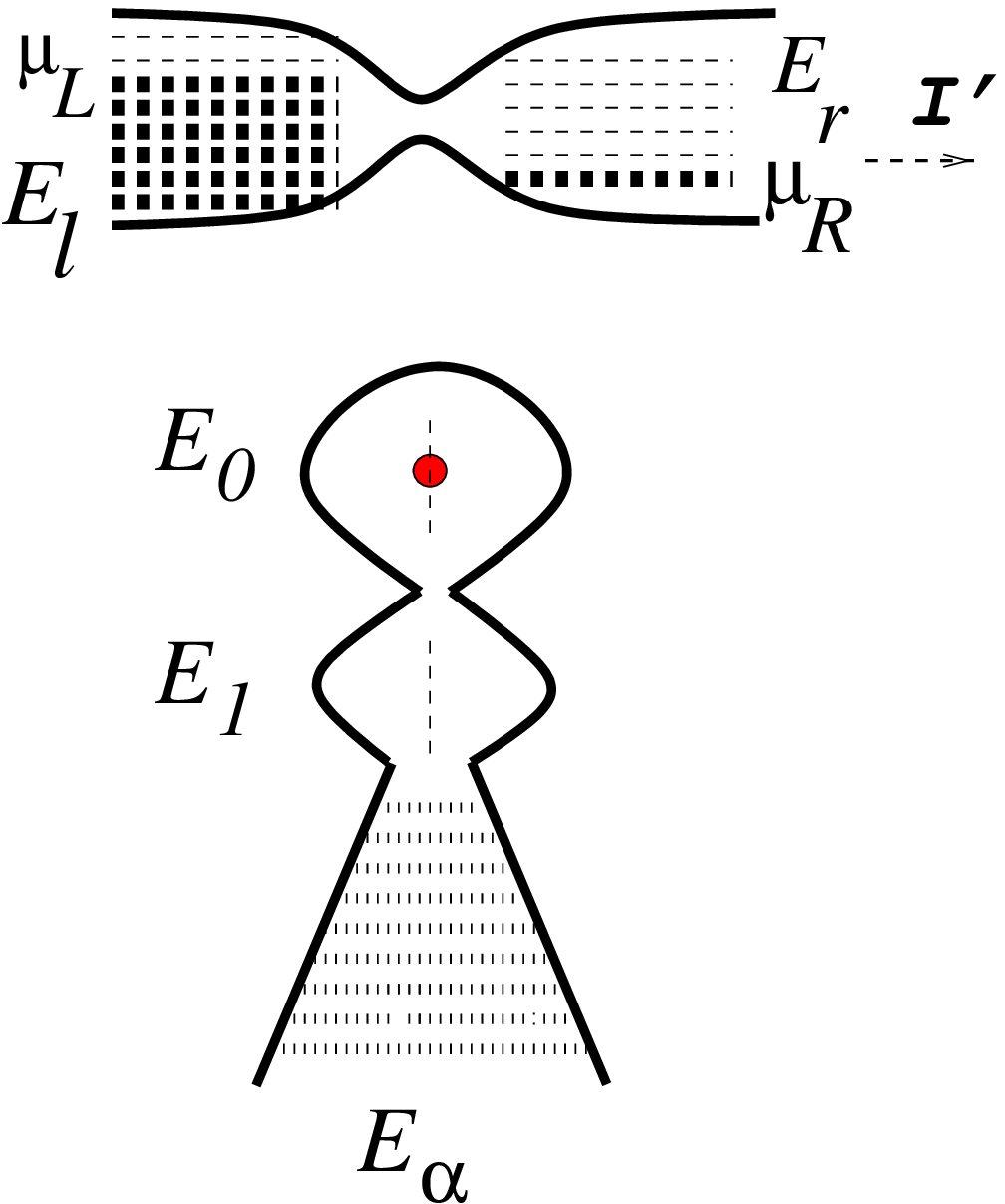}
\end{center}
{\begin{small}
{\bf Fig.~6.} A point-contact detector near a quantum dot coupled
with a resonance cavity.
\end{small}}
\end{minipage} \\ \\ 

Using the same treatment as in the previous section and summing over  
the states $n$ of the detector one arrives at the following rate
equations for the electron density matrix
$\sigma_{ij}(t)=\sum_n\sigma_{ij}^{(n)}(t)$,
\begin{subequations}
\label{dd}
\begin{eqnarray}
\label{dda}
\dot{\sigma}_{00} & = & i\Omega_\alpha (\sigma_{01}-\sigma_{10})\\
\label{ddb}
\dot{\sigma}_{11} & = & -\Gamma_1\sigma_{11}
+i\Omega_\alpha (\sigma_{10}-\sigma_{01})\\
\label{ddc}
\dot{\sigma}_{01} & = & i\epsilon_{10}\sigma_{01}
+i\Omega_\alpha (\sigma_{00}-\sigma_{11})
-\frac{\Gamma_1+\Gamma_d}{2}\sigma_{01}\\
\label{ddd}
\dot{\sigma}_{\alpha\alpha} & = & i\Omega_\alpha(\sigma_{\alpha 0}
-\sigma_{0\alpha})\\
\label{dde}
\dot{\sigma}_{0\alpha} & = & i\epsilon_{\alpha 0}\sigma_{0\alpha}
+i\Omega_\alpha (\sigma_{00}-\sigma_{1\alpha})
-\frac{\Gamma_d}{2}\sigma_{0\alpha }\\
\label{ddf}
\dot{\sigma}_{1\alpha} & = & i\epsilon_{\alpha 1}\sigma_{1\alpha}
+i\Omega_\alpha (\sigma_{10}-\sigma_{0\alpha })
-\frac{\Gamma_1}{2}\sigma_{1\alpha }\ ,
\end{eqnarray} 
\end{subequations}
where the index ``1'' relates to the cavity state ($E_1$)
and $\Gamma_d=(\sqrt{D}-\sqrt{D'})^2$ is 
the decoherence rate generated by the detector.

Consider first the case of no measurement, $\Gamma_d=0$. Solving
Eqs.~(\ref{dd}) we find that the decay is not a pure
exponential one \cite{eg1}.  In particular, the  
probability of finding the electron in the initial state for small $t$ is
$\sigma_{00}(t)= 1-\Omega_\alpha^2t^2$, 
in contrast with Eq.~(\ref{Req1a}). This second-order
dependence of $\sigma_{00}$ on $t$ is due to the fact that decoupling it from the 
off-diagonal $\sigma_{01}$ results in a second-order differential 
equation. Note that the absence of a term linear in $t$ in 
$\sigma_{00}(t)$ is a prerequisite for the Zeno effect. However,
at large values of $t$ the decay becomes an exponential one, i.e.
\begin{equation}
\sigma_{00}(t)\simeq\exp \left (-\frac{4\Omega^2_\alpha}{\Gamma_1}\ t\right )
~~~~{\mbox {for}}~~~ t\gg 1/\Omega_\alpha\, .
\label{d2}  
\end{equation}

Consider now the case of measurement, i.e.~$\Gamma_d\not =0$.  
Solving Eqs.~(\ref{dd}) for $t\gg\Omega_\alpha^{-1}$ we
discover that the probability
of finding the electron inside the dot, $\sigma_{11}(t)$,
drops down exponentially as
\begin{equation}
\sigma_{00}(t)= \exp\left (-\frac{4(\Gamma_1+\Gamma_d)
    \Omega_\alpha^2}{4(E_1-E_0)^2+(\Gamma_1+\Gamma_d)^2}t\right )
~~~~{\mbox {for}}~~~ t\gg 1/\Omega_\alpha\, . 
\label{d3}
\end{equation}

Let us compare Eq.~(\ref{d3}) with  Eq.~(\ref{d2}). We see 
that the decay rate decreases with $\Gamma_d$ (Zeno effect) in the presence of the
detector, but only for aligned levels,
$|E_0- E_1|\ll\Gamma_1+\Gamma_d$.
If, however, the levels $E_0$ and $E_1$ are not aligned,
$|E_0- E_1|\gg \Gamma_1+\Gamma_d$,
we find that the decay rate increases
(the anti-Zeno effect \cite{gur1,r6,az1,kof,eg1,facchi,evers}).
Such an increase of the decay rate due to the measurement is shown in
Fig.~7a for $E_1-E_0=10\Omega_\alpha$. However, for
very short times we always observe the decrease of the decay rate 
i.e. the Zeno effect even for misaligned levels, as shown in
Fig.~7b \cite{eg1}. Thus the anti-Zeno effect discussed recently
in the literature represents an increase in the ``average'' transition
rate. For small enough $t$, however, no anti-Zeno effect can be
found. 
\vskip1cm
\begin{minipage}{13cm}
\begin{center}
\includegraphics[width=12cm]{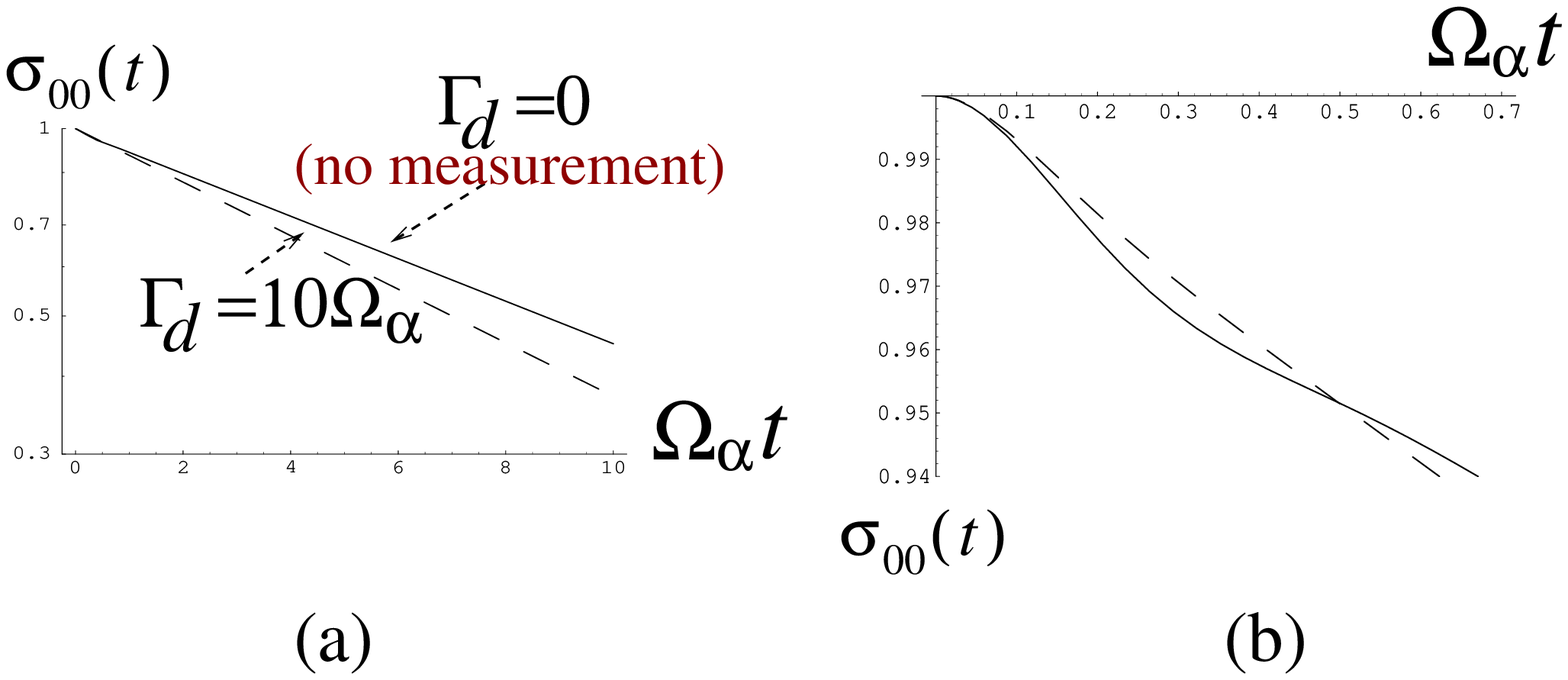}
\end{center}
{\begin{small}
{\bf Fig.~7.} (a) The probability of finding the electron inside the quantum dot
at the level $E_0$ where $E_1-E_0=10\Omega_\alpha$. The solid line
corresponds to $\Gamma_d=0$ (no measurement) while 
the dashed line, which displays the anti-Zeno effect, corresponds  
to $\Gamma_d=10\Omega_\alpha$. 
(b) The same for small $t$, where the dashed line
displays the Zeno effect.
\end{small}}
\end{minipage} \\ \\ 

The Zeno and anti-Zeno effects described above can be  
interpreted in terms of broadening of the level $E_0$ induced by  
the detector. One expects that this broadening would always 
lead to spreading of the energy distribution. On the other hand,  
its influence on the decay rate depends 
whether the levels $E_0$ and $E_1$ are in resonance or not. If
$E_0= E_1$ the broadening of the level $E_0$ destroys
the resonant-tunneling condition, so that the decay into continuum 
slows down. If on the other hand $E_0\not =E_1$, 
the same broadening would effectively diminish the levels displacement
(c.f. with \cite{luis}). As a result, the decay rate should increase. 
Yet, such qualitative arguments do not work at very short times,
since the decay rate slows down even for $E_0\not =E_1$, as in
Fig.~7b., except in the case of a flat density of states,
for which the decay rate is not affected by measurement. 

\section{Origin of the wave-function collapse.}

We demonstrated in this paper that the inclusion of a measurement device
in the Schr\"odinger equation made it possible to describe the measurement
process without explicit
use of the projection postulate (the wave-function collapse).
Even the Zeno effect is described in terms of the
decoherence generated by a macroscopic detector. Thus
one might assume that wave-function collapse is a redundant
assumption and can be avoided by including the detector in   
the Schr\"odinger equation for an entire system. Yet this is not the case: 
wave-function collapse is the indispensable component
of Quantum Mechanics.

Let us explain this point again taking the point-contact detector in Fig.~2 as an example.
We demonstrated above how tracing over the quantum dot subsystem
reduces the Schr\"odinger equation describing such a detector
to the classical rate equations Eq.~(\ref{aa11}),
\begin{equation}
\dot P_n(t) = -D\, P_n(t)
+D\, P_{n-1}(t)\, ,
\label{e1}
\end{equation}
where $P_n(t)\equiv\sigma^{(n,n)}(t)$ is the probability of
finding $n$ electrons in the right reservoir by time~$t$. 
Solving these equations for the initial conditions
$P_n(0)=\delta_{n0}$, we obtain a Poisson distribution
for $P_n(t)$, Eq.~(\ref{b4}). Consider for simplicity the case
of $t\gg 1/D$. Then $P_n(t)$ can be written as 
\begin{equation}
P_n(t)\simeq \frac{1}{\sqrt{2\pi Dt}}
\exp\left [-\frac{(Dt-n)^2}{2Dt}\right ].
\label{e2}
\end{equation}

Let us assume that the detector displays $N_1$ electrons at $t=t_1$, 
and the corresponding information is directly available to the observer. One can
ask whether such an information affects  
the distribution function $P_n(t)$, Eq.~(\ref{e2}). Simple
arguments show that it does. Indeed, Eq.~(\ref{e1}) represents 
classical rate equation and therefore it obeys 
Bayes principle \cite{bye}, as any probabilistic description. 
This implies that one has to solve Eq.~(\ref{e1}) with
the new initial condition, determined by the information obtained by
the observer. One obtains \cite{gur3} 
\begin{equation}
P_n(t)\simeq \frac{1}{\sqrt{2\pi D(t-t_1)}}
\exp\left [-\frac{(Dt-n+\Delta N)^2}{2D(t-t_1)}\right ]
\label{e3}
\end{equation}
where $\Delta N=N_1-Dt_1$. Obviously, this distribution is
different from that given by Eq.~(\ref{e2}): it has a narrower 
width, but the same group velocity. This result is not
surprising, since the probabilistic description of classical
systems is not a complete one. The measurement improves our knowledge
of the system, so the statistical uncertainty diminishes. 

The above arguments must also be applicable to Eq.~(\ref{e1}) considered 
as a pure quantum mechanical equation. Indeed, Eq.~(\ref{e1}) has been obtained
directly from the Schr\"odinger equation in the limit of high bias
voltage and without any use of the Markov-type anzatz. Thus
the Schr\"odinger evolution must be subject to 
Bayes principle too. As a matter of fact, Bayes principle, extended to
the off-diagonal density-matrix elements, is essentially
equivalent to wave function collapse \cite{kor}. 

Despite the importance of the Bayes principle in any probabilistic
description, it does not appear in standard calculations of 
Quantum Mechanics, since the latter does not predict individual events but
only ensemble averages of observables and their 
correlations. This allowed us to avoid  explicit use of
the projection postulate in the above evaluations of
the detector average current and the measured 
electron distributions.

Now it would be interesting to compare 
our result for the Zeno effect with alternative predictions
involving the projection postulate. On first sight some of our
results contradict such predictions. For instance, 
we predict that continuous measurement does not affect the
decay rate for a flat density of final states, contrary to the
the projection postulate arguments leading to the 
Zeno effect, Eq.(\ref{a3}).
However, for a flat density of states and infinite
reservoirs, the expansion (\ref{a1}) is not applicable due to
a discontinuity in the derivative $\dot P_0(t)$ at $t=0$. In this case
the probability of survival drops linearly with $t$ for small
$t$, and no Zeno effect is expected from the
projection postulate argument. For a nonuniform distribution,
however, we obtain a decrease of the decay rate for
small $t$, as shown in Fig.~7b.

In any case, such a comparison of standard quantum mechanical
calculations with those involving the projection postulate 
is far from being completed. For a proper understanding
of the measurement process we need to extend our quantum description
of the detector to a chain of measurement devices representing the
von Neumann hierarchy \cite{neu} (a system ``measured'' by another
system etc). Only then one can properly investigate a possible
dynamical role of the projection postulate 
in Quantum Mechanics.  We believe that our 
quantum rate equations, described in this paper, represent the  
proper tool for the realization of this program. 

\section{Summary} 

In this paper we have proposed Bloch-type rate equations
as a very useful approach to the quantum mechanical treatment
of measurement devices. These quantum rate equations were 
derived from the microscopic Schr\"odinger equation
without any stochastic assumptions.

First we applied this approach to a quantum mechanical treatment
of the point-contact detector. The latter represented a generic example of a
measurement device for the continuous monitoring of
an unstable system. We found that the transition 
to the classical regime of the detector takes place due to decoupling of the
nondiagonal density-matrix elements from the equations of motion for the diagonal.
The latter does not require the vanishing of these terms.

Then we used the same approach for a description of a larger system
consisting of an observed electron which escapes into continuum
together with the
point-contact detector. The decoherence mechanism is clearly
displayed in the resulting rate equations, where    
the corresponding decoherence rate is determined by the averaged
detector outcome. 

With respect to Zeno effect, we found that the measurement would
not affect the decay rate of an unstable system, providing that
the density of final states is a flat one and the
reservoir is infinite. If this is not the case, we
predict either Zeno or anti-Zeno effects,
except for the short-time behavior
where only the Zeno effect is found.

All our results were obtained without explicit use of the
projection postulate. Nevertheless, the latter cannot be discarded in
Quantum Mechanics, as we demonstrated using the example of
point-contact detector evolution. We have also shown that
our quantum mechanical predictions for the decay rate measurements
are not in contradiction with the projection postulate
argument, although a more detailed analysis is needed.

\end{document}